\documentclass[twocolumn,preprintnumbers,
endnote,nofootinbib,prl]{revtex4}
\usepackage{graphicx}

\usepackage[hypertex]{xcolor}

\newcommand{\vev}[1]{\langle {#1} \rangle}
\newcommand{\lsim}{\lesssim}
\newcommand{\gsim}{\gtrsim}

\newcommand{\eq}[1]{Eq.~(\ref{#1})}

\newcommand{\ord}[1]{\mathcal{O}{(#1)}}
\newcommand{\beq}{\begin{equation}}
\newcommand{\eeq}{\end{equation}}
\newcommand{\bea}{\begin{eqnarray}}
\newcommand{\eea}{\end{eqnarray}}

\newcommand{\mP}{M_{\rm P}}

\newcommand{\ug}{U(1)_g}
\newcommand{\ugp}{U(1)_g'}
\newcommand{\eS}{e^{-S}}

\begin{document}

\pagestyle{plain}

\title{\boldmath Gravitational Interactions and Neutrino Masses}

\author{Hooman Davoudiasl
\footnote{email: hooman@bnl.gov}
}

\affiliation{\vskip0.25cm
High Energy Theory Group, Physics Department\\ Brookhaven National Laboratory,
Upton, NY 11973, USA}


\begin{abstract}

We describe a scenario where the smallness of neutrino masses is related to a global 
symmetry that is only violated by quantum gravitational effects.  The coupling of neutrinos to gauge singlet right-handed fermions is attributed to symmetry preserving gravitational operators suppressed by the Planck mass, in this framework.  The proposed scenario leads to axion particles that decay into neutrinos, which could be probed through cosmological measurements and may help explain the Hubble parameter tension.  Depending on the details of the implementation, the scenario could provide axion dark matter candidates.       

\end{abstract} \maketitle

\underline{\it Introduction:} The two outstanding mysteries of particle physics and cosmology - the origin of 
small neutrino masses and the nature of 
dark matter (DM) - provide the most compelling phenomenological evidence for new 
physics.  While neutrinos are well-established ingredients of the Standard Model (SM), 
we do not know what type of physics provides the DM content of the Universe - about 25\% of its total energy budget \cite{PDG}.  

The smallness of neutrino masses, $m_\nu \lsim 0.1$~eV, could be a manifestation of ultraviolet (UV) 
physics, or it could be explained by a tiny dimensionless coupling $\sim \ord{10^{-12}}$.  A popular example of UV physics that could explain why neutrino masses are small is the seesaw mechanism with ultra-heavy right-handed neutrinos, leading to left-haded Majorana states at low energies \cite{seesaw}.  The heavy right-handed neutrinos could be as heavy as $\sim 10^{14}$~GeV.  
Larger masses would lead to non-perturbative couplings and are not generally considered. This  mechanism predicts low energy lepton number violation 
manifested as rare neutrinoless double beta decays, suppressed by small 
Majorana masses $\lsim 0.1$~eV.  Despite its theoretical appeal, 
the seesaw picture of small $m_\nu$ is quite challenging to verify experimentally and may not yield to direct confirmation. 

In principle, one could imagine that some global symmetry forbids neutrino masses at the renormalizable level.  Then, if this symmetry is very weakly broken we could end up with tiny $m_\nu$\footnote{An Abelian gauged symmetry in this spirit was suggested to describe quark mass hierarchies long ago, in Ref.~\cite{Froggatt:1978nt}.}.  
However, it is generally expected that gravitational effects 
lead to violations of global symmetries.  A macroscopic version of this expectation posits that a black hole destroys global charges and is fully described by mass, spin, and gauge charges.  In this work, we will consider a scenario where the smallness of $m_\nu$ is protected by a global symmetry which is explicitly broken only by non-perturbative gravitational effects, as will be described below. 

Recent work on implications 
of non-perturbative gravitational processes on low energy effective theories can be found in 
Refs.~\cite{Fichet:2019ugl,Daus:2020vtf,Calmet:2019frv}.  Ref.~\cite{Dvali:2016uhn} considers  neutrino condensation and masses that are gravitationally induced; for work along this direction using a different approach see Ref.~\cite{Barenboim:2019fmj}.  In Ref.~\cite{Dudas:2020sbq}, a scenario connecting DM and sterile neutrinos through gravitational interactions has been examined.  For prior work where generation of heavy right-handed 
neutrino masses from string theory instanton effects was considered see Ref.~\cite{Ibanez:2006da}.  
We will give a more concrete description of our scenario below.  However, we will first provide some  
clarifying comments.    

\underline{\it Caveats:} Before going further, we would like to clarify a few points.  First, the gravitational effects of interest here can only be fully determined in a consistent theory of quantum gravity, which is still under investigation.  Nonetheless, 
string theory seems to contain all the necessary ingredients for such a framework and many qualitative results can be gleaned from its possible structures.  From a general relativistic point of view, semi-classical studies of wormholes \cite{Abbott:1989jw,Kallosh:1995hi} and black holes also offer such insights.  While we make no pretense that this work represents an {\it ab initio} treatment, we will use ideas and results inspired 
by the above well-motivated approaches to argue for a qualitative picture of how neutrino masses may be low energy manifestations of Planck scale gravitational processes.  Obviously, we will not present a definitive model here, but instead we will aim to illustrate the 
general phenomena that could arise, and their possible signals, in this picture .

\underline{\it Organizing Principles:}  In this work, we will entertain the possibility that 
there is a global $U(1)_g$ symmetry that demands 
$m_\nu = 0$.  We will assume, consistent with the above considerations, that the $\ug$ symmetry is only violated by operators that are suppressed by non-perturbative ``gravitational instantons.''  Operators that do not violate $\ug$ are present in the low energy 
effective theory, possibly suppressed by powers of 
Planck mass $\mP\approx 1.2 \times 10^{19}$~GeV.  This setup gives rise to axions whose mass is 
generated by gravitational effects that explicitly break the global symmetry.  
This is analogous to the well-known Peccei-Quinn mechanism \cite{Peccei:1977hh} that was proposed to resolve the strong CP puzzle and yields an axion whose mass is generated by 
QCD instantons \cite{Weinberg:1977ma,Wilczek:1977pj}.  

Our approach has elements in common with the Majoron model \cite{Chikashige:1980ui}, in that it involves a global symmetry that leads to axions.  However, the Majoron models address the generation of typically large masses for right handed singlet neutrinos that provide the basis for the seesaw mechanism.  These models lead to light {\it Majorana} masses for the SM neutrinos.  In our work, as will be detailed below, we  will only consider generating {\it Dirac} masses, which require unusually small Yukawa couplings whose  
explanation lends itself well to a Planck-suppressed mechanism.  {\it Hence, specific models that we will  describe later can be falsified if neutrinoless double beta decay is observed.}   

The Majoron model has also been considered in the context of gravitational global 
symmetry violation in Ref.~\cite{Rothstein:1992rh}.  Our approach differs from that 
of Ref.~\cite{Rothstein:1992rh} in that we do not allow gravitational symmetry violations, 
unless they are mediated 
by ``instanton'' effects, making such violations exponentially suppressed.  We adopt the view that these instantons represent tunneling between neighboring vacua with different $\ug$ charges, in a similar fashion that 
non-perturbative electroweak processes allow transitions among vacua with different $b+l$ charges,  
where $b$ and $l$ are baryon and lepton numbers, respectively \cite{tHooft:1976rip}.  

In our treatment, operators that are only suppressed by powers of $\mP$ are presumably generated by perturbative and non-perturbative gravitational effects, but they do not   result in global charge violation (a similar approach was adopted in Ref.~\cite{Calmet:2019frv}, motivated by the results of Ref.~\cite{Kallosh:1995hi}).  However, since gravity acts universally on all types of particles, 
we use these gravitational operators to connect fields that do not share any other 
type of interactions and may not be from the same physical sector.  In particular,   
to generate Dirac masses for neutrinos, ``right-handed neutrinos'' with no gauge charges are required.  We suggest that it is plausible that such fermions are not part of the SM sector, but 
could couple to the SM fields through gravitational interactions suppressed by powers of 
$\mP$; if these operators do not violate global charges, there is no additional ``instanton'' suppression. 

In what follows, we will assume that new physics, 
such as supersymmetry, which is required to have a consistent 
UV theory of quantum gravity appears only at or close to $\mP$.  This has the advantage that in minimal implementations of our proposal, Planck suppressed operators that set the effective 
Yukawa couplings for neutrinos provide fairly definite predictions for the required scale of spontaneous global symmetry breaking.

Having laid out the organizing principles of our work, we will next provide more specific details 
for choices of parameters.

\underline{\it Instanton Action:} We will take the aforementioned gravitational instanton effects to  correspond to an action $S$.  While the size of this action depends on the 
details of the underlying spacetime geometry and the quantum theory of gravity, it has been argued that a typical string theory inspired size for $S$ is given by 
\beq
S\sim \frac{2 \pi}{\alpha_G}\,,
\label{S}
\eeq 
where $\alpha_G \sim 1/25$ is roughly the grand unified gauge coupling 
\cite{Svrcek:2006yi,Hui:2016ltb}.  
Absent a strong 
motivation for a particular value, for the illustrative purposes of our work here we will generally  
assume that \cite{Hui:2016ltb} 
\beq
1/30 \lsim \alpha_G \lsim 1/20 \;\; \Rightarrow \;\; \eS \sim 10^{-82}-10^{-55}. 
\label{eSval}
\eeq
As we will show later, the above choice yields numerically interesting results that demonstrate the utility of our scenario, while corresponding to UV motivated values.

Let us denote SM singlet fermions, often called right-handed neutrinos, by $\nu_R$; we postulate that these fermions have charge $Q_g(\nu_R)$ under $\ug$, assumed to be respected at the classical and renormalizable level.  As mentiond before, we will take the general view that since these states are not charged under any SM interactions, they can reasonably be expected to be from an entirely different sector and only couple to the SM neutrinos through ``gravitational interactions.'' 

If the dim-4 ``Dirac mass'' 
term $H^*\bar L \nu_R$ is forbidden by $\ug$, it will lead to zero neutrino masses; here $H$ is the Higgs doublet field with  vacuum expectation value (vev) $\vev{H}\approx 174$~GeV and $L$ is an SM lepton doublet.  However, as mentioned earlier, it is generally expected that non-perturbative gravitational effects would not respect $\ug$.  Yet such violations of the associated charge would be 
exponentially suppressed by $e^{-\Delta Q_g S}$, where $\Delta Q_g$ is the net magnitude of the charge of the operator.  
 
As a first attempt, it seems natural to assume that we only need to have $Q_g(\nu_R) = 1$, with all SM fields uncharged 
under $\ug$ (we will later show why this minimal setup would not yield acceptable values of 
$m_\nu$).  Therefore, 
we could have a Dirac mass term $\eS H^*\bar L \nu_R$ in the low energy effective theory.  However, this interaction 
will lead to negligibly tiny masses $\lsim 10^{-44}$~eV for neutrinos, given the reference values in \eq{eSval}.  Note that a ``Majorana'' mass term for $\nu_R$ of the form $ e^{-2 S}\,\mP \nu_R \nu_R$ could be generated through gravitational effects, but it would be extremely small $\lsim 10^{-91}$~eV.  For comparison, the inverse size of the visible Universe, given by the present day Hubble parameter ${\cal H}_0 \sim 10^{-33}$~eV, is enormously larger.  Hence, for all intents and purposes both types of neutrino masses are zero at the renormalizable level.  

Next, we will consider a 
minimal model, dubbed ``Model I,'' that accommodates viable values of $m_\nu$ and leads to potentially  observable cosmological signals. 

\underline{\it Model I:} 
Since the gravitationally generated dim-4 interactions do not yield the inferred values of $m_\nu\sim 
0.1$~eV, we need to consider other contributions from higher dimension operators.  Note that operators of the form $(H L)^2/\mP$ would generate Majorana masses for neutrinos that are about 5 orders of magnitude too small.  Therefore, we are led to consider additional fields that allow forming $\ug$ neutral operators, to avoid severe suppressions from instanton effects.     

Let us introduce a scalar $\Phi$ with $\ug$ charge $Q_g(\Phi)$.  We will attempt to generate acceptable ``Dirac masses'' $m_\nu$.  Hence, to avoid further 
suppressions, we need to make sure that any induced Majorana masses for $\nu_R$ satisfy 
$m_R \ll m_\nu$.  Generally speaking, we then need to ensure that operators of the type 
$\Phi^n \nu_R \nu_R$, with $n\geq 1$, are sufficiently suppressed.  We will hence choose a set of charges that will 
make this possible and also lead to operators that can provide the right size of $m_\nu$.   
Our choice for the purposes of illustration here will be $(Q_g(\Phi),Q_g(L), Q_g(\nu_R))=(1, -2, -3)$, with all other fields uncharged under $\ug$.  Note that this charge assignment is presumably not unique, but we will show that it could lead to interesting results.  In what follows, we will refer to this choice as ``Model I.'' 

With the above charges, we can write down the following dim-5 operator 
\beq
O_5 \sim \frac{\Phi H^*\bar L \nu_R}{\mP}\,,
\label{dim5}
\eeq
which has $\Delta Q_g = 0$ and hence can be generated by gravitational effects unsuppressed by 
instanton effects.  If the vev of $\Phi$ is non-zero, $\vev{\Phi} = \phi_0/\sqrt{2}$, we will then get neutrino Yukawa couplings to the Higgs  
$y_\nu$ of the size
\beq
y_\nu \sim \frac{\vev{\Phi}}{\mP}.
\label{ynu}
\eeq 
To get the correct mass for the neutrinos, we need $y_\nu \vev{H} \sim 0.1$~eV.  This requires $y_\nu \sim 10^{-12}$ and hence $\vev{\Phi} \sim 10^{7}$~GeV.

To generate $\vev{\Phi}\neq 0$, we consider the potential 
\beq
V(\Phi) = - m^2 \Phi^\dagger \Phi + \lambda (\Phi^\dagger \Phi)^2\,,
\label{Vphi}
\eeq
where $m$ is the mass parameter of $\Phi$ and $\lambda$ is its $\ord{1}$ self-coupling constant.  
Given the above considerations, we expect a heavy scalar 
$\phi$ of similar mass, $m_\phi \sim \vev{\Phi}$, upon spontaneous breaking of $\ug$.  However, if the 
$\ug$ is fully respected we would also end up with a massless ``Goldstone'' boson or axion $a$ 
in the low energy effective theory.  The vev $\vev{\Phi}$ is then identified with the 
decay constant of $a$.  In passing, we note that if the $\ug$ breaking entails a first order phase transition, it could lead to primordial gravitational waves.  However, the above scale is a factor 
of $\sim 10^2$ beyond the sensitivity of future ground-based gravitational wave observatories \cite{Croon:2018kqn}.   Assuming a lower gravitational scale than $\mP$, perhaps corresponding to 
a more fundamental description, could possibly bring the requisite symmetry breaking scale $\vev{\Phi}$ within the reach of those experiments.

As discussed earlier, we expect gravitational effects to 
violate $\ug$ through the action of non-perturbative instantons and thus to  
generate a potential for $a$, given by \cite{Svrcek:2006yi}  
\beq
V_a \sim - \eS \mP^4 \, \cos \frac{a}{\phi_0}\,.
\label{Va}
\eeq
The above yields a mass for the axion 
\beq
m_a^2 \sim \eS \frac{\mP^4}{\phi_0^2}\,,
\label{ma2}
\eeq
which for our choice of parameters yields 
\beq
10^{-10}~\text{GeV} \lsim m_a \lsim 3\times 10^3~\text{GeV} \quad (\text{Model I}).
\label{maVals}
\eeq

Let us parametrize $\Phi$ as 
\beq
\Phi = \frac{\phi + \phi_0}{\sqrt{2}} \,e^{i a/\phi_0}.  
\label{Phi}
\eeq
Using \eq{dim5}, we find that the coupling of $a$ to Dirac neutrinos $\nu$ is given by 
\beq
g_a \, a\,  \bar \nu \gamma_5 \nu =  
\frac{\vev{H}}{\sqrt{2} \mP} \, a\,  \bar \nu \gamma_5 \nu = 
\frac{m_\nu}{\phi_0} \, a\,  \bar \nu \gamma_5 \nu\,,
\label{anunu}
\eeq
which is - as expected for an axion - proportional to neutrino masses and suppressed by the axion 
decay constant.  We then find $g_a \sim 10^{-17}$ for the above model.  The lifetime $\tau_a$ of $a$ from decay into neutrinos is given by 
\beq
\tau = \frac{8 \pi}{g_a^2 \, m_a} \sim 10^{13}~\text{s} \; \; 
\left(\frac{\text{20 MeV}}{m_a}\right)
\left(\frac{10^{-17}}{g_a}\right)^2\,.
\label{tau}
\eeq
For comparison, the age of the Universe is 
$t_U\sim 4 \times 10^{17}~\text{s} \sim(2\times 10^{-42}~\text{GeV})^{-1}$ and the the cosmic microwave background (CMB) era roughly corresponds to $t_{CMB}\sim 10^{13}$~s.    
Given the possibility of a cosmologically long lifetime for the axion, it is interesting to consider it as a  
possible signature of the above model, as we will discuss next.

The decay of DM could leave an observable imprint on the evolution of the Universe.  
In Ref.~\cite{Poulin:2016nat}, this general possibility was considered and its effects on CMB and matter power spectra were examined.  These authors find that in the long lifetime regime, which roughly corresponds to $\tau \gsim t_U$, the fraction $f$ of DM that decays at a rate $\Gamma$ is bounded by \cite{Poulin:2016nat} 
\beq
f \, \Gamma < 15.9 \times 10^{-3}~\text{Gyr}^{-1}\quad (95\% \; \text {CL}).  
\label{fGamma}
\eeq
Using \eq{tau}, we can recast the above bound as 
\beq
f \,m_a< \frac{8.9\times 10^{-42}}{g_a^2}~\text{GeV}.
\label{fma}
\eeq
In the intermediate regime, corresponding roughly to $t_{CMB} \lsim \tau \lsim t_U$, based on the analysis of Ref.~\cite{Poulin:2016nat}, we use the typical bound $f \lsim 0.038$.

The initial energy density stored in $a$ is of order $m_a^2 a_i^2/2$, where $a_i$ 
is the initial amplitude of the axion oscillations, commencing when $m_a \approx 3 {\cal H}$, 
where ${\cal H}= (c_* \, g_*)^{1/2} T^2/\mP $ - with $T$ the temperature, $c_*\equiv (2\pi)^3/90$, and $g_*$ the relativistic degrees of freedom -  is the Hubble parameter during the radiation dominated era.  The oscillating modulus energy density drops with the expansion of the Universe like 
that of matter, that is like $T^3$.  We interpret $f$ to be the ratio of the axion energy density to that of cosmic DM at 
matter-radiation equality  marked by $T_{\rm eq} \sim$~eV; 
for $f=1$ the axion is assumed to constitute all DM.  Thus, demanding the axion energy density  redshift to $\sim f T_{\rm eq}^4$ at $T\sim T_{\rm eq}$, we find
\beq
f \approx \frac{a_i^2}{2} \left(\frac{9 c_*\,g_*}{\mP^2}\right)^{3/4}\left(\frac{\sqrt{m_a}}{T_{\rm eq}}\right).
\label{f}
\eeq


The above formula, however, should be used with care, since at sufficiently large values of $m_a$ the 
temperature at which $a$ starts its oscillation is large compared to $\phi_0$.  
Assuming that the field $\Phi$ is initially in thermal equilibrium, one would expect that it 
gets a vev after a phase transition at $T\sim \phi_0$.  Hence, for sufficiently large $m_a$, the condition $m_a \approx 3 {\cal H}$ would correspond to temperatures where the symmetry $\ug$ is typically unbroken and there is no axion.  As a representative range of parameters, we consider $m_a\in [10^{-3}, 20]$~MeV; this range would correspond to the onset of axion oscillation temperatures $T\lsim 10^8$~GeV, with the axion typically expected to be present (due to spontaneously broken symmetry).  We note that potential electroweak gauge boson couplings proportional to anomalies are not required and could in principle be set to zero;  this may require further assignment of charges.  Hence, thermal production of $a$ does not pose an impediment to implementation of our scenario.

\begin{figure}
\includegraphics[width=0.45\textwidth]{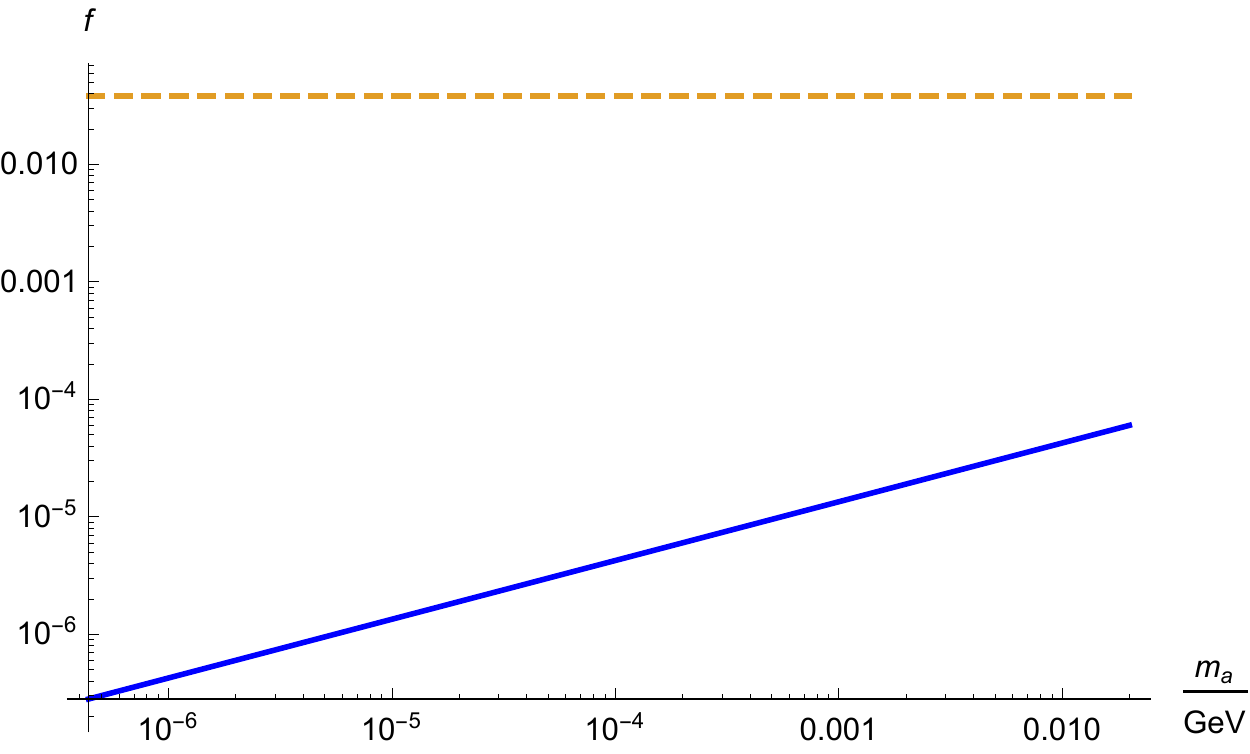}
\caption{Unstable fraction $f$ of dark matter versus $m_a$ in Model I from \eq{f}.  Values above the dashed line are excluded, 
corresponding to the 95\% C.L. limits from Ref.~\cite{Poulin:2016nat} in the $t_{CMB}\lsim \tau \lsim t_U$ regime (see the text for details).}
\label{fvsma}
\end{figure}

For $m_a\in [10^{-3}, 20]$~MeV, assuming the maximum amplitude $a_i=\phi_0$ and $g_*\sim 100$, we present the 
values of $f$ from \eq{f} versus $m_a$ in Fig.~\ref{fvsma}, shown as the solid line.   The horizontal dashed line is the intermediate lifetime ($t_{CMB} \lsim \tau \lsim t_U$) 95\% C.L. bound $f\lsim 0.038$, from Ref.~\cite{Poulin:2016nat}.  Values of $f$ above this line are excluded.  Constraints beyond the intermediate lifetime bound, corresponding to $m_a \gsim 20$~MeV ($\tau\lsim t_{CMB}$) are weaker \cite{Poulin:2016nat}.  Also, the long lifetime constraints ($\tau \gsim t_U$) are not very constraining and are not shown.  For $m_a \gsim 20$~MeV, as long as the reheat temperature is 
much larger than $\phi_0$, we generally expect that the axion starts its oscillation  
only upon spontaneous symmetry breaking and hence $f$ grows with $m_a^2$ in this mass range.  
For $m_a \sim 2$~GeV, corresponding to $\tau \sim 0.01 t_{CMB}$, we roughly get $f\sim 1$, which 
suggests above this mass one perturbs standard cosmology, since the new unstable component 
of DM starts to be significant and the parameters are likely not viable \cite{Poulin:2016nat}.

Our mechanism could potentially provide a 
resolution of a persistent tension between values of present time 
Hubble parameter ${\cal H}_0$ determined from the CMB \cite{Aghanim:2018eyx} and local  
\cite{Riess:2011yx,Riess:2018byc} measurements, with the latter yielding a result that is a few standard deviations larger than that obtained from the former.  We note that this tension could be a result of underestimated or unknown systematic uncertainties, however it has persisted for some time and its significance has been at an interesting level (recent work in Ref.~\cite{Riess:2019cxk} suggests that it now stands at $4.4 \sigma$).  Hence, it is reasonable to entertain 
the possibility that it could be due to new physics.  

One of the proposed resolutions of the above ${\cal H}_0$ tension postulates 
late time DM decay into dark particles or radiation \cite{Berezhiani:2015yta,Enqvist:2015ara,Anchordoqui:2015lqa,Pandey:2019plg,Vattis:2019efj}; see also Ref.~\cite{Poulin:2016nat}.  
Such resolutions of the ${\cal H}_0$ tension could require that only a sub-dominant component of DM decay by the present epoch; see, for example, Ref.~\cite{Berezhiani:2015yta}.  We would then need a cosmologically stable component, that we will not specify here, to account for the DM observed today.  Axion decays in Model I would lead to a population of relativistic neutrinos that behave like 
dark radiation.  A more detailed study is required to examine whether our scenario could plausibly alleviate the Hubble parameter tension.  Nonetheless, given 
that the general features of a resolution are present in our proposal, let us elaborate on this 
possibility some more.

If axions make up a fraction $f$ of DM energy density, 
today's flux could be of order 
\beq
F_0^\nu\sim f \,\rho_{DM}/m_a, 
\label{Fnu}
\eeq
where $\rho_{DM} \sim 1.3\times 10^{-6}$~GeV cm$^{-3}$ 
is the cosmic value of DM energy density.   We are interested in DM decay after the CMB era, corresponding to $t_{CMB}\sim 10^{13}$~s.  From \eq{tau}, $\tau \gsim t_{CMB}$ requires $m_a\lsim 20$~MeV.  Using \eq{f}, we then have $F_0^\nu \gsim 100$~cm$^{-2}$ s$^{-1}$.  To 
see if this flux is detectable we need to know its typical energy at the present time. 

During the matter dominated era, corresponding to $t \gsim t_{CMB}$,  
we have $t \sim R^{3/2}$, with $R$ the cosmic expansion scale factor.  
If the decay takes  place at time $t_d \sim \tau$ (instantaneous approximation), then the energy of the neutrinos at present time $E_0^\nu$ is roughly given by
\beq
E_0^\nu \sim \frac{m_a}{2} \left(\frac{\tau}{t_U}\right)^{2/3},
\label{E0nu}
\eeq 
where the energy of decay final state neutrinos is assumed to be $\sim m_a/2$, {\it i.e.} of order the cosmologically unstable DM mass, which we have identified with $m_a$.  Hence, for $m_a \lsim 20$~MeV, we find $E_0^\nu\lsim 10$~keV.  This energy is small enough that it presents a challenge to detection, which typically requires $\ord{\text{MeV}}$ energies.

\underline{\it Model II:} 
To explore further possibilities of the gravitational neutrino mass generation scenario  
proposed here, let us consider 
a simple extension of the above setup.  Though this comes at the expense of minimality, it 
would lead to potentially interesting and broader options for phenomenology.   We will call  
this extension ``Model II.''  Here, we propose to 
expand the model by another global symmetry $\ugp$.  From a UV (string theory) point of view, 
one expects a multitude of such symmetries, see {\it e.g.} Ref.~\cite{Svrcek:2006yi}.  In principle, 
$\ug$ and $\ugp$ symmetries would be violated by separate instantons 
of action $S$ and $S'$, respectively.  We will assume that $S = S'$, as distinct numerical values are 
not necessary for our illustrative examples, given the broad range considered in \eq{eSval}.   

We will also 
assume that there is an additional scalar $\Phi'$.  The $\ug$ charge assignments of the fields are as follows: $(Q_g(\Phi), Q_g(\Phi'), Q_g(L), Q_g(\nu_R)) = (1, 0, q+1, q)$ and the corresponding 
$\ugp$ assignments are $(0,1,0,-1)$.   Therefore, we can write down the following gravitationally 
mediated dim-6 operator
\beq
O_6\sim \frac{\Phi \Phi' H^*\bar L \nu_R}{\mP^2}\,,
\label{dim6}
\eeq
which, in order to generate the correct size for $m_\nu$ requires 
\beq
\frac{\vev{\Phi}\vev{\Phi'}}{\mP^2}\sim 10^{-12}.
\label{PhiPhip}
\eeq

As before, any dim-4 Dirac masses for neutrinos would be exponentially suppressed by instantons.  We will assume the same instanton processes would lead to violations of $\ug$ and $\ugp$ symmetries.  
Also, Majorana masses for $\nu_R$ would dominantly originate from operators of the from 
\beq
\frac{\vev{\Phi}^{2q} \vev{\Phi'}^2 \nu_R\nu_R}{\mP^{2q+1}}.
\label{MnuR}
\eeq
For $q=2$, we find a Majorana mass $\sim 4\times 10^{-17}$~eV which is $\ll m_\nu$ and hence we could 
take $m_\nu$ to be a Dirac mass generated from the $O_6$ operator in \eq{dim6}, to excellent accuracy. 

Let $\vev{\Phi}=\phi_0/\sqrt{2}$ and $\vev{\Phi'}=\phi_0'/\sqrt{2}$; we will denote the axions 
associated with these fields by $a$ and $a'$, respectively.  Let us choose 
$\phi_0 = 10^9$~GeV and $\phi_0'=10^{17}$~GeV, as illustrative examples.  By analogy with the discussion of Model I 
and \eq{ma2}, we find 
\beq
10^{-12}~\text{GeV} \lsim m_a \lsim 30~\text{GeV} \quad (\text{Model II})
\label{maValsII}
\eeq
and 
\beq
10^{-20}~\text{GeV} \lsim m_{a'} \lsim 3\times 10^{-7}~\text{GeV} \quad (\text{Model II}).
\label{mapValsII}
\eeq
One could show that the couplings of the $a$ and $a'$ axions to neutrinos in Model II are, as expected, 
$g_a=m_\nu/\phi_0 \sim 10^{-19}$ and $g_{a'}=m_\nu/\phi_0'\sim 10^{-27}$, respectively. 

\begin{figure}
\includegraphics[width=0.45\textwidth]{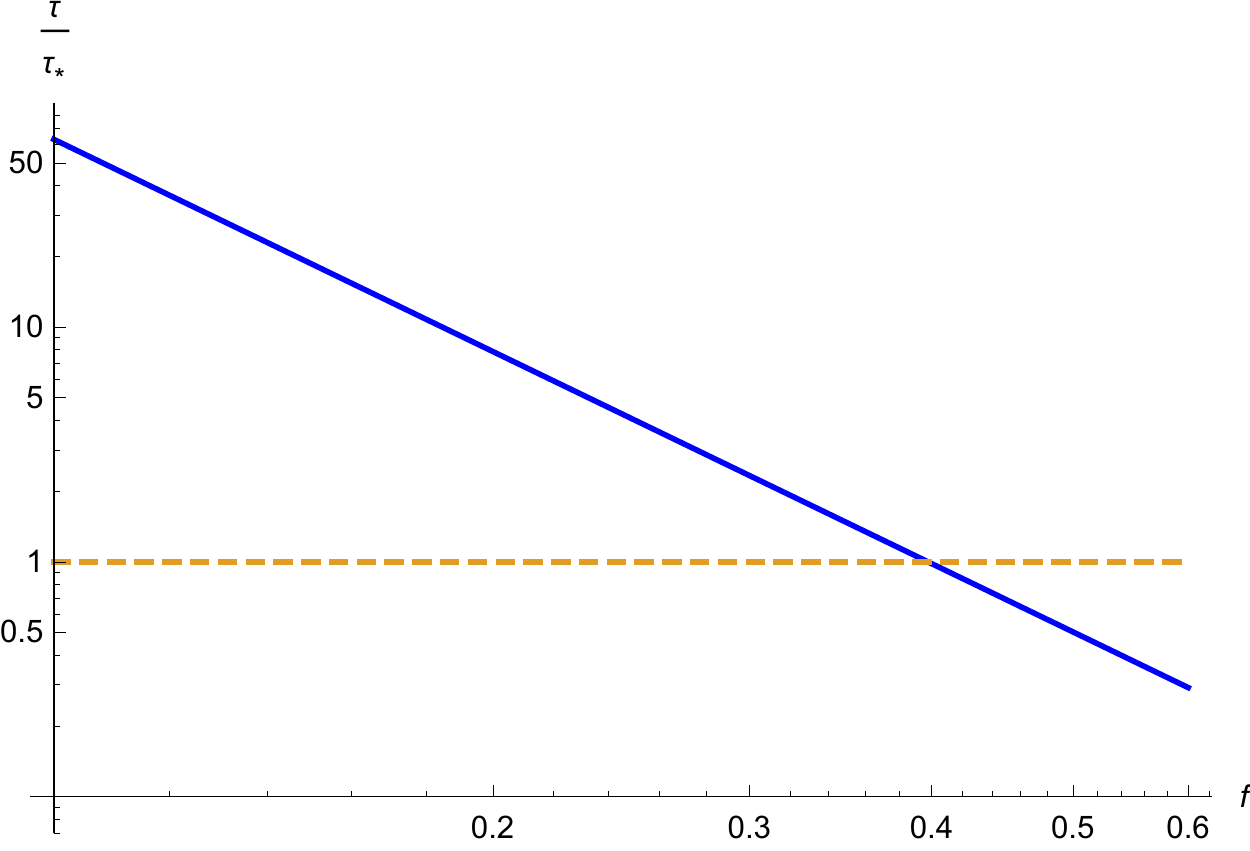}
 \caption{Lifetime, in units of $\tau_* = f \,2.0\times  10^{18}~\text{s}$, for the unstable DM fraction $f$ of axion $a$ in Model II, assuming $a_i = \phi_0$ (see the text for details).  Values below the dashed line, 
corresponding to the 95\% C.L. limit (for $\tau \gsim t_U$) from Ref.~\cite{Poulin:2016nat}, are excluded.}
 \label{tauvsf}
\end{figure}

The bound in \eq{fGamma}, relevant to the regime $\tau \gsim t_U$, is equivalent to 
\beq
\tau > f \,2.0\times  10^{18}~\text{s}\,.
\label{taubound}
\eeq
Using Eqs.~(\ref{tau}) and (\ref{f}), and assuming $a_i = \phi_0$, in 
Fig.~\ref{tauvsf} we have plotted the values of 
the axion $a$ lifetime in units of $\tau_*\equiv f \,2.0\times  10^{18}~\text{s}$, 
as a function of $f$.  The model parameters are allowed by the cosmological constraints above the horizontal dashed line.  Hence, Model II can lead to a non-negligible fraction of unstable DM that decays on time scales comparable to or longer than $t_U$, and thus a potentially detectable flux of neutrinos from 
$a\to \bar \nu \nu$.   Let us estimate the flux of these neutrinos.  

We will assume that there is a spherical distribution of DM particles centered around the Earth, with radius $D$.  One can then show that the flux $F$ of neutrinos from the decay of DM arriving at Earth is given by
\beq
F \approx \frac{D\, f \rho}{m\, \tau}\,,
\label{flux}
\eeq
where $\rho$ is the DM energy density.  In the Galactic neighborhood of the Solar System, we have 
$\rho_S \approx 0.3$~GeV cm$^{-3}$ \cite{Tanabashi:2018oca}, say, for $D \sim 0.5$~kpc.  
For $m_a\sim 5$~MeV and $a_i=\phi_0$, using \eq{f}, we find $f\sim 0.3$ and 
$\tau \sim 4 \times 10^{17}$~s.  Based on this set of possible parameters, we see that the flux of  
neutrinos with energy $E_\nu \sim 2.5$~MeV will be given by 
$F \sim 10^5$~cm $^{-2}$ s$^{-1}$, including a mixture of flavors of both neutrinos and anti-neutrinos.  Interestingly, this is not far from the level of ``geo-neutrino'' flux  that has been observed by both KamLAND and Boerxino collaborations \cite{Araki:2005qa,Bellini:2010hy}.  For example, the KamLAND result for the flux of $\bar \nu_e$ is $3.4 ^{+0.8}_{-0.8}\times 10^6$~$\text{cm}^{-2} \text{s}^{-1}$.  We see that current the uncertainty in this measurement  \cite{Gando:2013nba} is an order of magnitude above the neutrino flux from axion decays; similar conclusions apply to the 
Borexino results \cite{Agostini:2019dbs}. 
In principle, more precise measurements of geo-neutrino  flux,  together with more accurate geological models, could probe parts of the parameter space of Model II.   (For constraints on neutrino flux from DM Majoron decays, using extraterrestrial anti-neutrino 
searches by Boexino \cite{Bellini:2010gn} and KamLAND \cite{Collaboration:2011jza}, see Ref.~\cite{Garcia-Cely:2017oco}.  These constraints would allow the rough sample reference values  adopted above, for non-degenerate neutrinos.)

\begin{figure}
\includegraphics[width=0.45\textwidth]{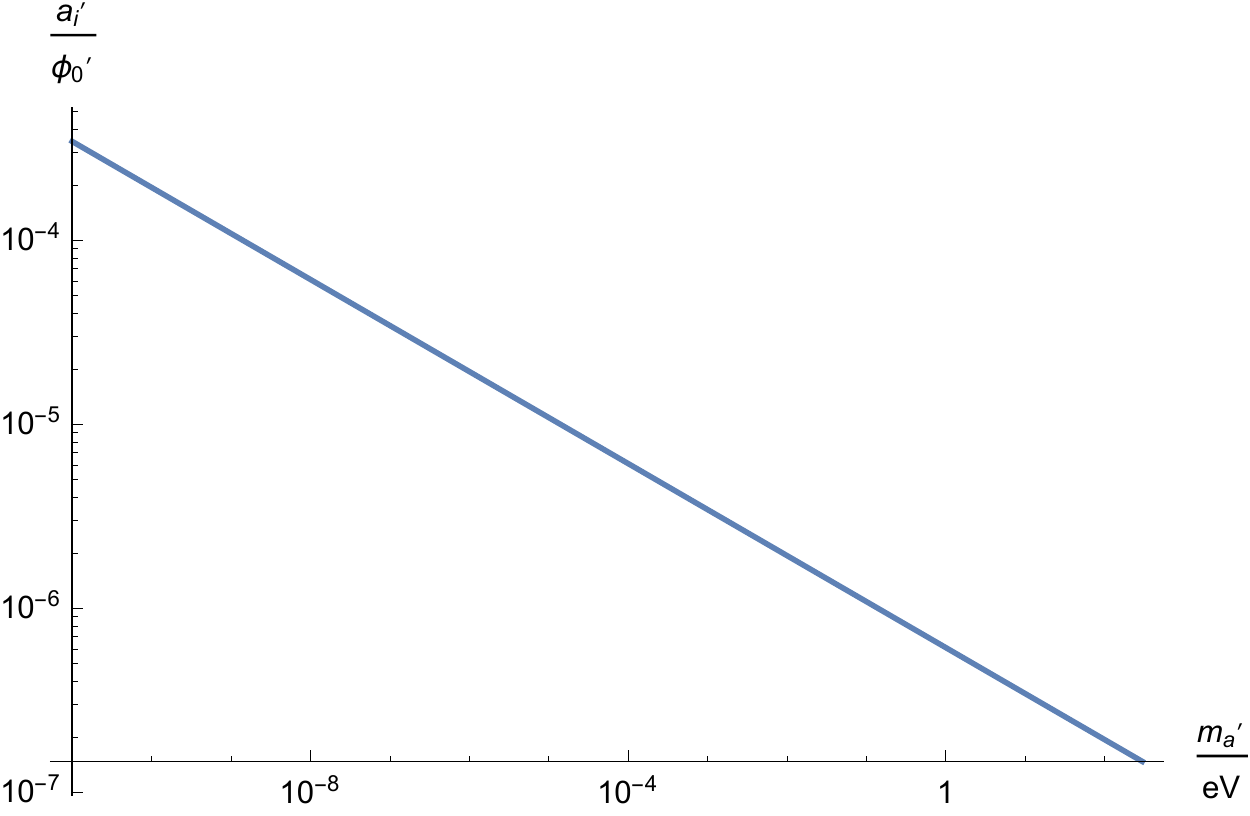}
 \caption{Initial amplitude of oscillations $a_i'$ in units of $\phi_0'$, versus $m_a'$ in Model II, assuming that the axion $a'$ constitutes all DM (see the text for more details).}
 \label{aipvsmap}
\end{figure}

In Fig.~\ref{aipvsmap}, we have presented the initial misalignment, in units of $\phi_0'$, 
needed for the axion $a'$ to 
constitute all DM, that is for $f=1$.  Note that the lifetime of $a'$ is much larger than $t_U$, 
by $\ord{10^{20}}$ or more, over the entire range of masses in \eq{mapValsII}.  Since over the 
reference range of $m_{a'}$ the misalignment required is less than unity, all values can represent 
potentially viable DM candidates.  For the lower end of the range, corresponding to 
$m_{a'}\gsim 10^{-11}$~eV, such DM may be produced copiously by spinning 
solar mass black holes \cite{Penrose:1969pc,YBZ}, which can be probed by gravitational wave measurements \cite{Arvanitaki:2009fg,Arvanitaki:2014wva}.

\underline{\it Summary:} 
To summarize, we have proposed that the small masses of neutrinos may be a hint for a global symmetry that requires them to vanish.  Such a symmetry is expected to be violated by exponentially suppressed non-perturbative gravitational ``instanton'' effects.  Our approach allows for symmetry preserving gravitational interactions among various fields, suppressed by powers of Planck scale but without instanton suppression.  In this view, ``right-handed'' neutrinos which may not be part of the SM sector could couple to it and generate Dirac masses for neutrinos.   Since we also require that the symmetry be spontaneously broken, our scenario leads to appearance of light axions, which generically couple 
to neutrinos and can decay into them.  As the axions could constitute a fraction or all of  
dark matter, we could expect interesting imprints of this scenario on 
cosmological evolution, as discussed in this work.  Simple extensions of the basic model can accommodate cosmologically stable dark matter axions, 
as well as an unstable axion population that could potentially lead to an observable neutrino flux.  We 
pointed out that the decay of this sub-dominant component into neutrinos could help alleviate the current tension between local and cosmological determinations of the Hubble parameter.

\acknowledgments

We thank Djuna Croon, Peter Denton, Julian Heeck, and Jure Zupan for comments and discussions.  This work is supported by the United States Department of Energy under Grant Contract DE-SC0012704.  




\end{document}